\documentclass[onecolumn,11pt]{elsarticle}

\usepackage{lineno,hyperref}
\modulolinenumbers[25]

\begin{document}

\begin{frontmatter}

\title{Skyrme fluid in anisotropic Universe. }
\author{ Rikpratik Sengupta$^a$, B C Paul$^b\footnote{$^*$Corresponding author.\\
{\it E-mail addresses:} rikpratik.sengupta@gmail.com (RS), bcpaul@associates.iucaa.in (BCP), prasenjit071083@gmail.com (PP) }$, Prasenjit Paul$^c$.}

\address{$^a$Department of Physics, Aliah University, Kolkata 700 160, West Bengal, India.\\
         $^b$Department of Physics, University of North Bengal, Siliguri 734 013, West Bengal, India.\\
         $^c$Department of Physics, Govt. College of Engineering and Ceramic Technology, West Bengal, India.}

\begin{abstract}
Cosmological solutions are obtained in an anisotropic Kantowski-Sachs (KS)  and Bianchi Type-I universes considering  a cosmological constant with Skyrme fluid. Interestingly, the solutions obtained here in both the KS and Bianchi-I anisotropic universes  are found isotropize at late time due to the presence of the Skyrme fluid even in the absence of $\Lambda$ term or any inflationary mechanism involving the inflaton field. A comparative study of both the  anisotropic cosmological models are carried out here and found that
	 Bianchi-I universe admits oscillatory solutions for a given matter configuration. We also note that the emergent universe model can be  obtained  with  Skyrme fluid. The  anisotropy, deceleration and jerk parameters have been studied along with the linear perturbative stability to explore efficacy of the model. Both the cosmological models are stable in the absence of cosmological constant besides their compatibility with observational  data. Thus, we claim Skyrme fluid a possible source for isotropization of an anisotropic universe $via$  accelerated expansion, which is capable of reproducing some observed features of the  universe.
\end{abstract}

\begin{keyword}
\texttt{Kantowski-Sachs, Bianchi-I, Skyrme 
fluid}
\end{keyword}

\end{frontmatter}

\linenumbers

\section{Introduction}

It is predicted from the cosmological observations that on a large scale the present universe is homogeneous and isotropic. However, immediately after the Big Bang, the scenario may have been different. It is believed that the early universe must have been highly anisotropic-the initial anisotropy being washed away in course of evolution of the universe with time \cite{Misner}. This suggests that the present state of the universe is not affected by it's initial conditions. Thus, space-times which are spatially homogeneous and anisotropic should provide a good description of the early universe. Also, the local anisotropies observed today in galaxies and superclusters motivate us to study space-times which are anisotropic. Anisotropic cosmological models are widely studied to study the possible effects of early universe anisotropy on present day observations.

The Bianchi Type-I cosmological model which is homogeneous with spatially flat sections but direction dependent expansion or contraction rate, are the simplest of such anisotropic cosmological models and have been studied in this context of present day observational signature in \cite{Huang,Chimento,Saha1,Saha2,Pradhan1,Pradhan2}. There are other types of Bianchi models as well, the whole set named after Luigi Bianchi who classified the relevant 3-dimensional spaces \cite{Bianchi}, which were first studied in the framework of the Einstein Field Equations (EFE) by \cite{Taub,Heckmann,Ellis}. Such anistoropic initial conditions have also been studied in Kantowski-Sachs(KS) cosmology\cite{Kantowski}. It was Collins \cite{Collins} who first described the global structure of KS space-time. There exists a class of KS solutions with a non-vanishing value of the Cosmological Constant in which the initial anisotropy washes out asymptotically \cite{Weber}. The isotropization of an anisotropic KS universe in the framework of Linde's chaotic inflationary scenario   \cite{Linde} studied by Paul {\it et. al.} \cite{Paul}. KS cosmologies have also been studied in details in a series of papers \cite{Dunsby1,Dunsby2,Dunsby3}.

Skyrme fields in the  background KS space-times have been considered in\cite{Canfora,Parisi}. It may be mentioned here that in particle physics the  skyrmion is considered as topologically stable field configuration in the framework of non-linear sigma models which was used for pions to model nucleons by Skyrme in 1962 \cite{Skyrme}. The model does not involve quarks and is an approximate low-energy effective theory of Quantum Chromodynamics (QCD). The Skyrme model is characterized by topological soliton solutions which are physically interpreted as a type of baryon termed $Skyrmions$ \cite{Kalbermann}. Thereafter, Skyrmions are considered in the field theoretical framework as a topological object which is of much interest in Solid State Physics and  String Theory.  An indirect evidence of the existence of Skyrmions have been reported in Bose-Einstein condensates and superconductors \cite{Khawaja,Baskaran}. In \cite{Canfora,Parisi} coupling of the Skyrme field with gravity has been considered taking the Skyrme fluid as a source. In cosmological context it generates   bounds on both the cosmological constant and the Skyrme coupling constant \cite{Canfora2}, and hence it is worth of further investigation to understand the evolution of the universe.

In the  paper we consider the EFE for anisotropic KS and Bianchi type-I space times with a non-zero cosmological constant taking the Skyrme fluid as a source. The evolution of the universe is studied for two different types of  anisotropic universes  considering different permissible values of the coupling parameters accommodating  different physical situations like domination of the Skyrme fluid. The paper has been organized as follows. In the following section we present the of the field equations of the Einstein-Skyrme system. In sections 3 and 4, cosmological solutions are obtained for KS and Bianchi I space-times respectively taking the Skyrme fluid as  a source with a cosmological constant $\Lambda$. In section 5 the physical interpretations of the models are discussed. The cosmological solutions are plotted graphically for comparative study of the cosmological models. The EFE for Bianchi I background geometry is represented in the appendix.

\section{The Einstein-Skyrme Action and the Field Equations}

The Einstein-Skyrme (ES) action with non-zero $\Lambda$ is given by 

\begin{equation}
	S=\frac{1}{16\pi G}\int d^4 x \sqrt{-g}(R-2\Lambda) + S_{Skyrme},
\end{equation}
where $ S_{Skyrme}$ is 
\begin{equation}
	S_{Skyrme}=\frac{\kappa}{2}\int d^4 x \sqrt{-g} \; \; Trace \left( \frac{R_{\mu \nu} R^{\mu \nu}}{2}+\frac{\lambda F_{\mu\nu}F^{\mu\nu}}{16} \right),
\end{equation}
where $\kappa=\frac{f_\pi^2}{4}$ and $\lambda=\frac{4}{e^2 f_\pi^2}$ are the Skyrmion coupling constants; ${f}_\pi$ and $e$ denote the pion decay constant and a dimensionless parameter for stable solitons, respectively.

The equation of motion (EOM) of a Skyrme field coupled to gravity is given by 
\begin{equation}
	\nabla^\mu R_{\mu \nu} +\frac{\lambda}{4} \; \nabla^{\mu} [R_{\mu \nu},F_{\mu\nu}]=0.
\end{equation}
The Einstein field equation (EFE) for a ES system with non-zero $\Lambda$ is given by
\begin{equation}
	G_{\mu\nu}+\Lambda g_{\mu\nu}=8\pi G \; T_{\mu\nu}^{Skyrme}.
\end{equation}
The energy-momentum tensor for the Skyrme fluid $T_{\mu\nu}^{Skyrme}$ is expressed as
\begin{equation}
	T_{\mu\nu}^{Skyrme}=(\rho+p_t)u_\mu u_\nu +p_t g_{\mu\nu} +(p_r-p_t)\chi_\mu \chi_\nu,
\end{equation}
where $u_\mu$ denotes the 4-velocity and $\chi^\mu=A^{-1} \delta_r^\mu$ is a unit space-like vector in the radial direction.

The above expression for the stress-energy tensor of the Skyrme field indicates the presence of  in-built anisotropy in the fluid as the pressures are different in the radial ($p_r$) and tangential  ($p_t$) directions. It is found that the  radial pressure is equal in magnitude but opposite in sign to the energy density. The fluid satisfies all the three energy conditions in General Relativity namely, weak energy condition (WEC), dominant energy condition (DEC) and strong energy condition (SEC). However in the presence of a $\Lambda$-term, it may or may not satisfy the Strong Energy Condition (SEC) which is  $p_t\geq \Lambda$ \cite{Parisi}.

In the above the gravitational constant $G$ multiplied by the Skyrme coupling constant $\kappa$ gives the effective gravitational constant $G_{eff}=G\; \kappa$. The parameter $8 \pi G \kappa$ may lie in the range  ( $0-1$) and the other coupling constant $\lambda$ is  approximately   of the order of $2 \, \times \,10^{-31} m^2$ \cite{Canfora2}.
In the anisotropic universes considered here the evolution  of the anisotropy parameter denoted by $\Gamma$ will be determined in the next sections.  

\section{Cosmological Solution for a KS universe with $\Lambda$ and Skyrme Fluid}

The KS space-time is described by the line element
\begin{equation}
	ds^2=-dt^2+A^2(t)dr^2+B^2(t)[d\theta^2+sin^2\theta d\phi^2]
\end{equation}
where $A(t) $ and $B(t)$ are the scale factors and the 4-dimensional coordinates $r$, $\theta$, $\phi$ and $t$.
The components of the $T_{\mu\nu}^{Skyrme}$ are given by \cite{Parisi}
\begin{equation}
	\rho=\frac{1}{B^2} \left(1+\frac{\lambda}{2B^2} \right).
\end{equation}
\begin{equation}
	p_r=-\rho,
\end{equation}
\begin{equation}
	p_t=\omega_t\rho=\frac{\lambda}{2B^4},
\end{equation}
where the equation of state (EOS) parameter corresponds to $\omega_t= \frac{\lambda}{\lambda+2B^2}$.
Considering the stress-energy tensor described by eqs. (7)-(9), the components of the EFE given by eq. (4) can be expressed as
\begin{equation}
	2 \frac{\dot{B}\dot{A}}{BA}+ \frac{1}{B^2}+ \frac{\dot{B}^2}{B^2}- \Lambda=8\pi G \left[\frac{\kappa}{B^2}\left(1+\frac{\lambda}{2B^2}\right) \right],
\end{equation}
\begin{equation}
	2 \frac{\ddot{B}}{B}+ \frac{1}{B^2}+ \frac{\dot{B}^2}{B^2}- \Lambda=8\pi G \left[\frac{\kappa}{B^2} \left(1+\frac{\lambda}{2B^2} \right) \right],
\end{equation}
\begin{equation}
	\frac{\dot{B}\dot{A}}{BA}+ \frac{\ddot{B}}{B}+ \frac{\ddot{A}}{A}- \Lambda=-8\pi G \left[\frac{\kappa\lambda}{2B^4} \right].
\end{equation}

\subsection{\bf KS Model-I}

Considering  a connection between the scale factors of the form $\dot{B}=KA$ we get
\begin{equation}
	\frac{\dot{A}}{A}=\frac{\ddot{B}}{\dot{B}}.
\end{equation}	
Using  the first two field eqs. (10) and (11) and  the above  relation we get
\begin{equation}
	2 \frac{\ddot{B}}{B}+ \frac{1}{B^2}+ \frac{\dot{B}^2}{B^2}- \Lambda= \frac{\alpha_1}{B^2}+\frac{\alpha_2}{B^4},
\end{equation}
where $\alpha_1=8\pi G\kappa$ and $\alpha_2=4 \pi G \kappa \lambda$.
The  EFE given by eq. (12) yields
\begin{equation}
	2 \frac{\ddot{B}}{B}+\frac{1}{\dot{B}}\frac{d^3 B}{dt^3}- \Lambda= -\frac{\alpha_2}{B^4}.
\end{equation}
Multiplying both sides of the above equation by $\dot{B}B^2$ and integrating the obtained differential equation(DE) we get

\begin{equation}
	\frac{\ddot{B}}{B}-\frac{\Lambda}{3}=\frac{\alpha_2}{B^4}+\frac{C}{B^3},
\end{equation}
where $C$ denotes the constant of integration.
Combining Equations(14) and (16), we get

\begin{equation}
	\frac{\dot{B}^2}{B^2}+ \frac{1-\alpha_1}{B^2}+ \frac{\alpha_2}{B^4}+ \frac{2C}{B^3}-\frac{\Lambda}{3}=0.
\end{equation}

We introduce a time-dependent perturbation $B(t) (1+\epsilon(t))$ to study the stability of the cosmological models. As we have seen from the EFE that the components of the stress-energy tensor are expressed in terms of the scale factor $B$ and we have also considered a relation between the two scale factors for obtaining the solutions to the EFE in  the Eq. (13) for the KS Model-I. So, involving perturbation terms up to the first power of $\epsilon$, we obtain from Eq. (17), a differential equation for the perturbation $\epsilon(t)$ as:

\begin{equation}
	\frac{\dot{B}}{B}\dot{\epsilon}-\left[\frac{1-\alpha_1}{B^2}+\frac{2\alpha_2}{B^4}\right]\epsilon= 0.
\end{equation} 

We now consider different cases below : 

\textbf{Case A}: $\alpha_2=0$, $C=0$ and $H_0^2=\frac{\Lambda}{3}$.
The eq. (17) reduces to
\begin{equation}
	\frac{\dot{B}^2}{B^2}+ \frac{1-\alpha_1}{B^2}= H_0^2.
\end{equation}
On integrating we obtain
\begin{equation}
	B=B_0 \; cosh \; (H_0 \; t),
\end{equation}
where $B_0=\frac{\sqrt{1-\alpha_1}}{H_0}$.
Thus the other scale factor is given by 
\begin{equation}
	A=\frac{\dot{B}}{K}=A_0 \; sinh \; (H_0 \;t),
\end{equation}
where $A_0=\frac{\sqrt{1-\alpha_1}}{\kappa}$. The two directional Hubble parameters are given by
\begin{equation}
	H_{1}=\frac{\dot{A}}{A}, \; \; H_{2}=\frac{\dot{B}}{B}
\end{equation}
The average Hubble parameter is given by
\begin{equation}
	H=\frac{H_{1}+2H_{2}}{3} 
\end{equation}
We define the variation of the directional Hubble parameters as
\begin{equation}
	\Delta H_{1}= H_{1}-H, \; \;  \Delta H_{2}= H_{2}-H
\end{equation}
The anisotropy parameter $\Gamma$ is defined as 
\begin{equation}
	\Gamma=\frac{1}{3}\bigg[\left(\frac{\Delta H_{1}}{H}\right)^2+2\left(\frac{\Delta H_{2}}{H}\right)^2\bigg]
\end{equation}
The anisotropy parameter for the above cosmological solution  yields
\begin{eqnarray}
	\Gamma&=&\frac{1}{3}\left[\bigg(\frac{H_1 - H}{H}\bigg)^2 + 2 \bigg(\frac{H_2 - H}{H}\bigg)^2 \right]\nonumber\\&=&\frac{2(coth^2 H_0t+tanh^2 H_0t-2)}{coth^2 H_0t+4 \;tanh^2 H_0t+4}.
\end{eqnarray}
The variation of the anisotropic parameter with time is shown  in Fig(\ref{fig:case3}).  It is evident that the initial anisotropy present in the early universe  washes out to zero finally in the late universe. 
The time-dependent perturbation has the form
\begin{equation}
	\epsilon(t)=C[\tanh H_0t]^{\frac{1-\alpha_1}{B_0^2H_0^2}}.
\end{equation}
For $\alpha_1=1$ it is constant but for $\alpha_1 <1$, it leads to a decreasing perturbation.

\textbf{Case B}: In this we consider $\alpha_2=0$, $C=0$, $H_0^2=\frac{\Lambda}{3}$ , however, the maximum value of the parameter $\alpha_1$ may be taken as $\alpha_1=1$. In this case on integrating eq. (17) the cosmological solution reduces to
\begin{equation}
	B=B_0 \; e^{H_0t},
\end{equation}
where $B_0$ is an integration constant and
\begin{equation}
	A=\frac{B_0 H_0}{\kappa} e^{H_0t}.
\end{equation}
Thus it corresponds to  de-Sitter  solution. As we get a vanishing anisotropy parameter ($\Gamma = 0$), the solution gives a universe which emerged from isotropic state. ALso, we see in this case that the perturbation turns out to be a constant and has no time-dependence at all.

\textbf{Case C}: In this we consider $\alpha_1=1$, $C=0$,  $H_0^2=\frac{\Lambda}{3}$  with a non zero $\alpha_2$. In this case the Skyrme fluid $dominates$ and integrating the eq. (17) we get 
\begin{equation}
	B=\frac{\alpha_2^\frac{1}{4}}{\sqrt{H_0}}\sqrt{cosh 2H_0t}.
\end{equation}
The other scale factor is given by
\begin{equation}
	A=\frac{\alpha_2^\frac{1}{4}\sqrt{H_0}}{K} \frac{sinh 2H_0t}{\sqrt{cosh 2H_0t}}.
\end{equation}
The anisotropy parameter becomes
\begin{equation}
	\Gamma=\frac{8 \; coth^2 H_0t+9 \; tanh^2 H_0t-16)}{4 \; coth^2 H_0t+4 \; tanh^2 H_0t+8},
\end{equation}
The time variation of the anisotropy parameter is shown  in Fig(\ref{fig:case6}). It is evident that an initial anisotropic universe transforms to an isotropic universe at late time. 
The perturbation on the scale factor is
\begin{equation}
	\epsilon(t)=\frac{C}{2H_0}[\tanh H_0t]^{2H_0}.
\end{equation}
 which decreases leading to a stable universe. 
 
\textbf{Case D}: In this case  Skyrme fluid is considered without cosmological constant  ($\Lambda$). Thus here we set $C=0$ in eq. (17) which leads to
\begin{equation}
	\frac{\dot{B}^2}{B^2}= \frac{\alpha_1-1}{B^2}-\frac{\alpha_2}{B^4}.
\end{equation}
Integrating the above differential equation we get the solution which is
\begin{equation}
	B=\sqrt{(\alpha_1-1)t^2+\frac{\alpha_2}{\alpha_1-1}}.
\end{equation}
\begin{equation}
	A=\frac{\alpha_1-1}{K}\frac{t}{\sqrt{(\alpha_1-1)t^2+\frac{\alpha_2}{\alpha_1-1}}},
\end{equation}
where  $\alpha_1 >1$, but if  $0 < \alpha_1  \leq 1$  it gives unphysical result.
The anisotropy parameter is given by,
\begin{equation}
	\Gamma=2\left[\frac{\frac{1}{t} - \frac{2t(\alpha_1 - 1)}{(\alpha_1 - 1)t^2 + \frac{\alpha_2}{(\alpha_1-1)}}}{\frac{1}{t} + \frac{t(\alpha_1 - 1)}{(\alpha_1 - 1)t^2 + \frac{\alpha_2}{(\alpha_1-1)}}}\right]^2,
\end{equation}
The time variation of the anisotropic parameter is shown  in Fig(\ref{fig:case7}).  It is evident that the initial anisotropy washes out which in a later epoch once again will transit to an anisotropic  universe.  However this particular solution is of no physical significance. Thus cosmological constant plays an important role for the evolution of the universe with skyrme fluid.

The time-dependent perturbation is found to evolve with time as
\begin{equation}
	\epsilon(t)=\frac{C\,t}{(\alpha_1-1)t^2+\alpha_1}.
\end{equation}
We plot the variation of the perturbation as a function of time  in Fig. 1, it is evident that a stable cosmological model can be obtained.

\begin{figure}[t]
	\centering
	\includegraphics[scale=0.4]{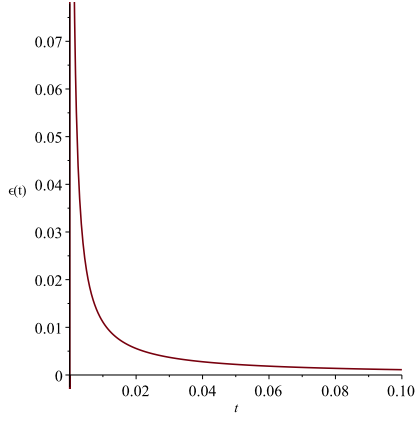}~\hspace*{-3.5cm}
	\vspace*{-5cm}
	\caption{variation of perturbation, $\epsilon(t)$ with time, $t$} \label{fig:case1}
\end{figure}

\subsection{\bf KS Model II}

Now we consider a simple case where $\dot{B}=0$ which leads to $\frac{1}{B}=H_0$ or $B=H_0^{-1}$.
The  eqs. (10) and (11) reduces to an identical equation which is given by
\begin{equation}
	\frac{1}{B^2}-\Lambda=\frac{\alpha_1}{B^2}+\frac{\alpha_2}{B^4}.
\end{equation}
The EFE given by eq. (12)  reduces to
\begin{equation}
	\frac{\ddot{A}}{A}-\Lambda=-\frac{\alpha_2}{B^4}.
\end{equation}
In this case the cosmological constant becomes
\begin{equation}
	\Lambda=[(1-\alpha_1)-\alpha_2H_0^2]H_0^2.
\end{equation}
Putting the value of $\Lambda$ back in eq. (12), we obtain
\begin{equation}
	\frac{\ddot{A}}{A}=[(1-\alpha_1)-2\alpha_2H_0^2]H_0^2.
\end{equation}
We note the following: \\
(i) For $H_0^2 = \frac{1-\alpha_1}{\alpha_2}$, $\ddot{A}=0$,  it gives rise to $linear$ evolution of $A$ with $t$.\\
(ii) For $H_0^2 > \frac{1-\alpha_1}{\alpha_2}$, $\ddot{A}<0$, one gets an $oscillatory$ universe solution.\\
(iii) For $H_0^2 < \frac{1-\alpha_1}{\alpha_2}$, $\ddot{A}>0$, one gets an  $expanding$ universe solution.

\section{Cosmological Solution for a Bianchi type-I universe}

The line element of Bianchi Type-I  anisotropic universe is given by 
\begin{equation}
	ds^2=dt^2-[R_i(t) \; dx_i]^2, i=1,2,3.
\end{equation}
where the anisotropic Hubble parameters, $H_i=\frac{\dot{R_i}}{R_i}$.
The average scale factor is given by $R=(R_1 R_2 R_3)^\frac{1}{3}$ and the average Hubble parameter is given by
\begin{equation}
	H=\frac{\dot{R}}{R}=\frac{d}{dt}(ln R)=\frac{1}{3}\frac{d}{dt}(ln V),
\end{equation}
where $V=R^3=R_1 R_2 R_3$ is the volume. We consider $R_1=A$, $R_2=R_3=B$ in this section, which implies $V=AB^2$ and $R=(AB^2)^\frac{1}{3}$.
The EFE are given by (refer to appendix)
\begin{equation}
	\frac{\dot{B}^2}{B^2}+ 2 \frac{\dot{A}\dot{B}}{AB}- \Lambda=8\pi G \; \left[\frac{ \kappa }{B^2}\left(1+\frac{\lambda}{2B^2} \right) \right].
\end{equation}
\begin{equation}
	\frac{\dot{B}^2}{B^2}+ 2 \frac{\ddot{B}}{B} -\Lambda=8\pi G \; \left[\frac{\kappa}{B^2} \left(1+\frac{\lambda}{2B^2} \right)\right].
\end{equation}
\begin{equation}
	\frac{\ddot{B}}{B}+ \frac{\ddot{A}}{A}+\frac{\dot{A}\dot{B}}{AB}- \Lambda=-8\pi G \; \left[\frac{ \kappa\lambda}{2B^4}\right].
\end{equation}

\subsection{\bf Bianchi-I cosmology}

We consider $\dot{B}=\kappa A $.
The eqs. (41) and (42) of the EFE reduce to a identical equation given by
\begin{equation}
	\frac{\dot{B}^2}{B^2}+2\frac{\ddot{B}}{B}-\Lambda= \frac{\alpha_1}{B^2}+\frac{\alpha_2}{B^4}.
\end{equation}
The eq. (43) reduces to
\begin{equation}
	2 \frac{\ddot{B}}{B}+ \frac{1}{\dot{B}}\frac{d^3 B}{dt^3}- \Lambda= -\frac{\alpha_2}{B^4}.
\end{equation}
Finally we get a differential equation for $B$ which is given by
\begin{equation}
	\frac{\dot{B}^2}{B^2}- \frac{\alpha_1}{B^2}+ \frac{\alpha_2}{B^4}+ \frac{2C}{B^3}-\frac{\Lambda}{3}=0.
\end{equation}
Similiar to KS Model-I, a time-dependent perturbation $\epsilon(t)$ can be introduced and the above equation involving perturbation corrections upto first power in $\epsilon$ and its derivative yields

\begin{equation}
	\frac{\dot{B}}{B}\dot{\epsilon}+\left[\frac{\alpha_1}{B^2}-\frac{2\alpha_2}{B^4}\right]\epsilon= 0.
\end{equation}

We now study different cases in the Bianchi I universe:

\textbf{Case A}: For $\alpha_2=C= 0$ and $H_0^2=\frac{\Lambda}{3}$, the above differential equation reduces to
\begin{equation}
	\frac{\dot{B}^2}{B^2}-\frac{\alpha}{B^2}=H_0^2.
\end{equation}
Integrating the above equation we obtain solution which is
\begin{equation}
	B=B_0 \; sinh H_0 \;t,
\end{equation}
where $B_0=\frac{\sqrt{\alpha_1}}{H_0}$ and
\begin{equation}
	A=A_0  \;cosh H_0 \;t,
\end{equation}
where $A_0=\frac{\sqrt{\alpha_1}}{\kappa}$.
The anisotropy parameter is given by,
\begin{equation}
	\Gamma=2\left[\frac{coth^2 H_0t + tanh^2 H_0t - 2}{4coth^2 H_0t + tanh^2 H_0t + 4}\right],
\end{equation}
The time variation of the anisotropic parameter is shown in Fig(\ref{fig:case3}). It is evident that the initial anisotropy in the universe  decreases and finally the anisotropic universe isotropized. 
The perturbation is given by
\begin{equation}
	\epsilon(t)=C[\tanh H_0t]^{\frac{-\alpha_1}{B_0^2H_0^2}}. 
\end{equation}

The evolution of the perturbation with time has been shown in Figure 2.

\begin{figure}[t]
	\centering
	\includegraphics[scale=0.4]{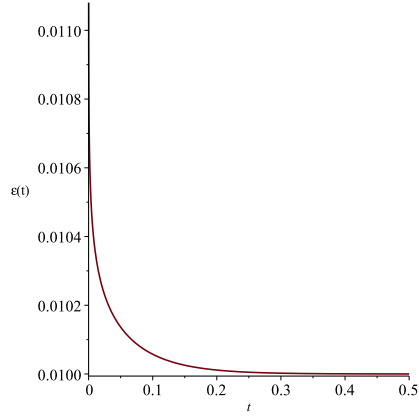}~\hspace*{-3.5cm}
	\vspace*{-5cm}
	\caption{variation of perturbation, $\epsilon(t)$ with time, $t$} \label{fig:case2}
\end{figure}

\textbf{Case B}: For $\alpha_1=1$, $\alpha_2=0=C$, $H_0^2=\frac{\Lambda}{3}$,  eq.(43) reduces to
\begin{equation}
	\frac{\dot{B}^2}{B^2}-\frac{1}{B^2}=H_0^2.
\end{equation}

\begin{figure}[t]
	\centering
	\includegraphics[scale=0.4]{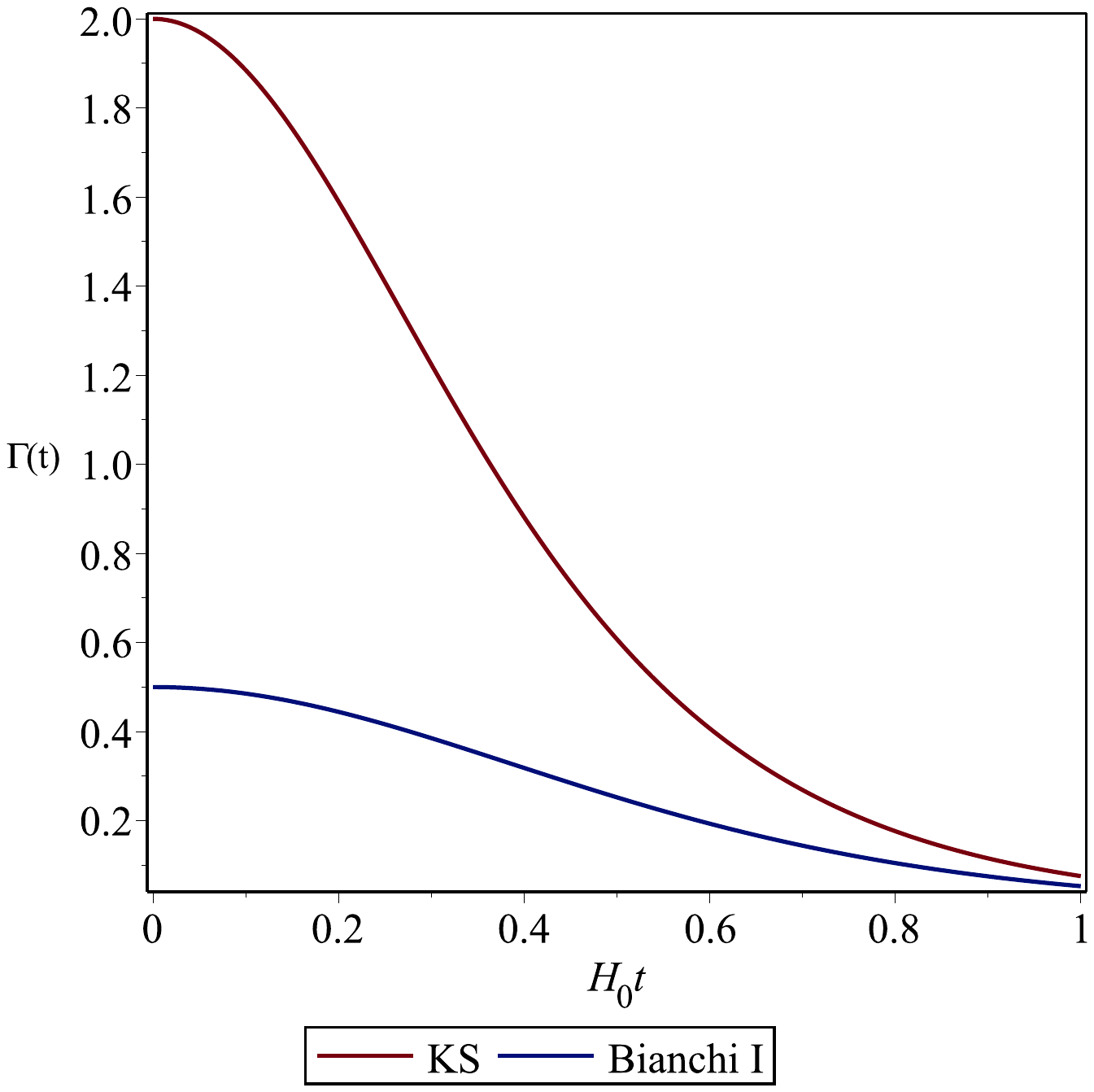}~\hspace*{-3.5cm}
	\vspace*{-5cm}
	\caption{variation of anisotropy parameter, $\Gamma(t)$ with time, $H_0t$} \label{fig:case3}
\end{figure}

\begin{figure}[t]
	\centering
	\includegraphics[scale=0.4]{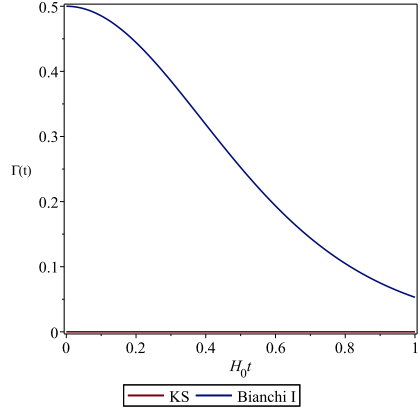}~\hspace*{-3.5cm}
	\vspace*{-5cm}
	\caption{variation of anisotropy parameter, $\Gamma(t)$ with time, $H_0t$} \label{fig:case4}
\end{figure}

The solution is given by
\begin{equation}
	B=B_0 \; sinh H_0 \;t,
\end{equation}
where $B_0=\frac{1}{H_0}$ and
\begin{equation}
	A=A_0 \; cosh H_0 \;t,
\end{equation}
where $A_0=\frac{1}{\kappa}$, which is similar to the Case A discussed above with $\alpha_1=1$. 
In this case the anisotropy parameter is given by
\begin{equation}
	\Gamma=2 \; \left[\frac{coth^2 H_0t + tanh^2 H_0t - 2}{4 \; coth^2 H_0t + tanh^2 H_0t + 4}\right],
\end{equation}
The time evolution of the anisotropic parameter  is shown  in Fig. (\ref{fig:case4}). A universe with initial anisotropy at a later epoch isotropizes. 
The perturbation takes the form
\begin{equation}
	\epsilon(t)=C[\tanh H_0t]^{\frac{-1}{B_0^2H_0^2}}.
\end{equation}
The temporal evolution of the perturbation is shown in Fig. 5.

\begin{figure}[t]
	\centering
	\includegraphics[scale=0.4]{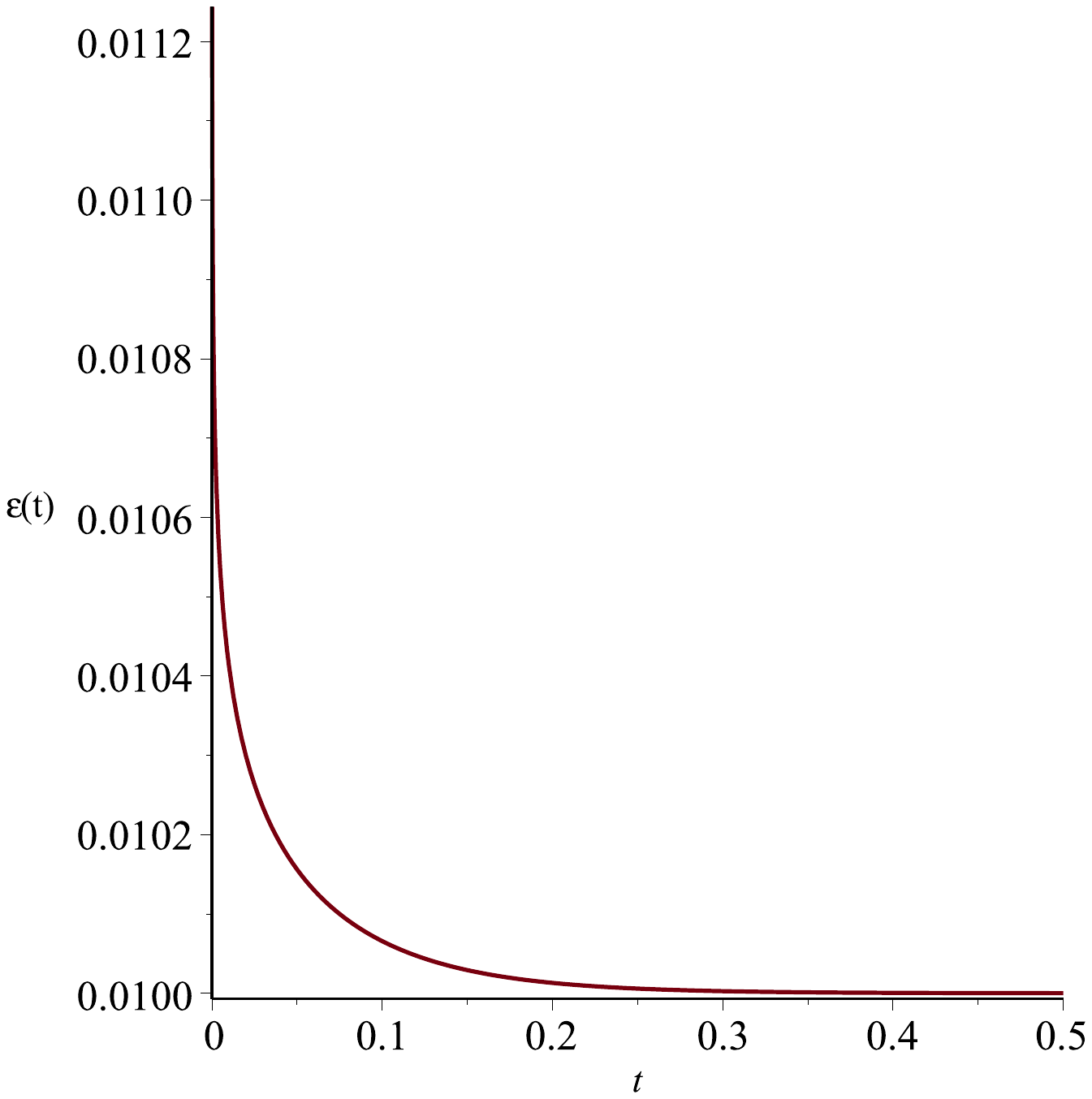}~\hspace*{-3.5cm}
	\vspace*{-5cm}
	\caption{variation of perturbation, $\epsilon(t)$ with time, $t$} \label{fig:case5}
\end{figure}

\textbf{Case C}: For $H_0^2=\frac{\Lambda}{3}$, $C=0$ and $\alpha_1=1$,  the Skyrme fluid $dominates$ as the coupling parameter $\kappa$ is set to a maximum value with non-zero  coupling parameter $\lambda$.
The differential equation for $B$ can be written as
\begin{equation}
	\frac{\dot{B}^2}{B^2}- \frac{1}{B^2}+ \frac{\alpha_2}{B^4}-\frac{\Lambda}{3}=0.
\end{equation}
On integrating once we obtain 
\begin{equation}
	B\dot{B}=\sqrt{H_0^2B^4+B^2-\alpha_2}.
\end{equation}
In the very early universe the value of $\Lambda$ was  very large, consequently one can  approximate $\frac{1}{H_0^2}\to 0$. The general solution  is given by 
\begin{equation}
	B=\sqrt{\frac{\sqrt{\alpha_2}}{H_0}cosh 2H_0t-\frac{1}{2H_0^2}}.
\end{equation}
\begin{equation}
	A=\frac{\frac{\sqrt{\alpha_2}}{K}sinh 2H_0t}{\sqrt{\frac{\sqrt{\alpha_2}}{H_0}cosh 2H_0t-\frac{1}{2H_0^2}}}.
\end{equation}
One interesting point to note is that if we consider  anti-deSitter universe with negative $\Lambda$, for this particular Bianchi-I solution, the scale factor $A$ will posses an additive term $\frac{1}{2H_0^2}$ in the denominator which can describe an emergent universe\cite{Mukherjee,Banerjee}. Thus an initial anisotropy with skyrme fluid admits an emergent universe which at a later epoch transits to an isotropic universe which is a new solution.
\begin{figure}[t]
	\centering
	\includegraphics[scale=0.4]{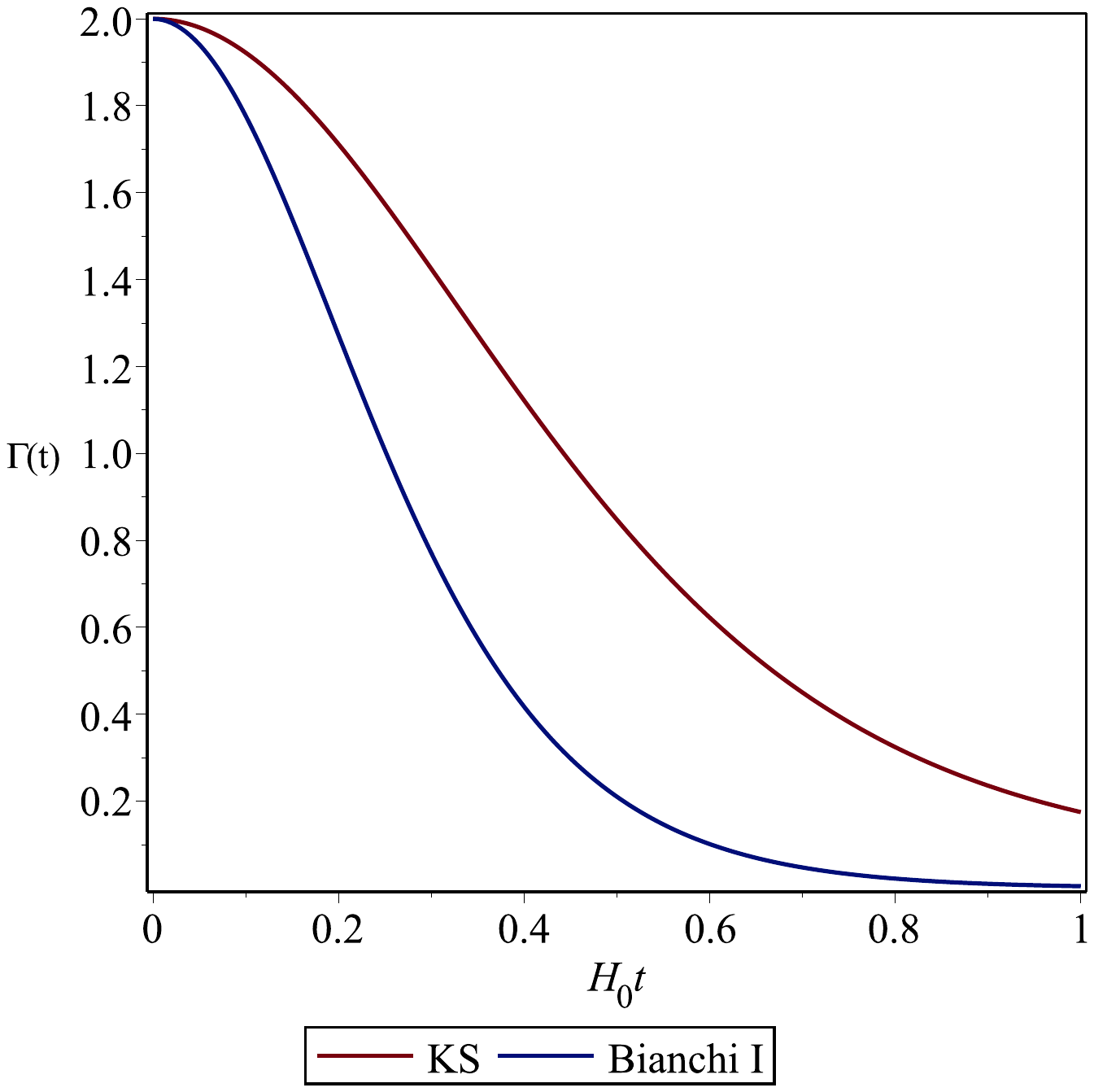}~\hspace*{-3.5cm}
	\vspace*{-5cm}
	\caption{variation of anisotropy parameter, $\Gamma(t)$ with time, $H_0t$} \label{fig:case6}
\end{figure}
	\begin{equation}
		\Gamma=\left[\frac{8H_0^2coth^2 2H_0t + \frac{8\alpha_2 sinh^2 2H_0t}{(\frac{\sqrt{\alpha_2} cosh 2H_0t}{H_0} - \frac{1}{2H_0^2})^2} - \frac{16H_0\sqrt{\alpha_2} cosh 2H_0t}{(\frac{\sqrt{\alpha_2} cosh 2H_0t}{H_0} - \frac{1}{2H_0^2})}}{4H_0^2coth^2 2H_0t + \frac{\alpha_2 sinh^2 2H_0t}{(\frac{\sqrt{\alpha_2} cosh 2H_0t}{H_0} - \frac{1}{2H_0^2})^2} + \frac{4H_0\sqrt{\alpha_2} cosh 2H_0t}{(\frac{\sqrt{\alpha_2} cosh 2H_0t}{H_0} - \frac{1}{2H_0^2})}}\right].
	\end{equation}
The time  variation of anisotropic parameter  is shown in Fig(\ref{fig:case6}). In this case the initial anisotropy also washes out.
The perturbation on the scale factor is given by 
\begin{eqnarray}
	\epsilon(t)&=&\frac{C}{2H_0}[\tanh H_0t]^{\big\{\frac{\frac{\alpha_2^\frac{3}{2}}{H_0}-\frac{1}{2H_0^2}-2\alpha_2}{\frac{\alpha_2}{H_0}-\frac{\sqrt{\alpha_2}}{2H_0^2}}\big\}}\big\{\frac{\sqrt{\alpha_2}}{2H_0^2}(\tanh H_0t)^2+\nonumber\\&&\frac{1}{2H_0^2}(\tanh H_0t)^2+\frac{\sqrt{\alpha_2}}{2H_0^2}-\frac{1}{2H_0^2}\big\}^\frac{8\alpha_2H_0^3}{4\alpha_2H_0^2-1}.
\end{eqnarray}

\textbf{Case D}: For $H_0^2=C=0$ without a  cosmological constant 
eq. (47) reduces to
\begin{equation}
	\frac{\dot{B}^2}{B^2}- \frac{\alpha_1}{B^2}+ \frac{\alpha_2}{B^4}=0.
\end{equation}

\begin{figure}[t]
	\centering
	\includegraphics[scale=0.4]{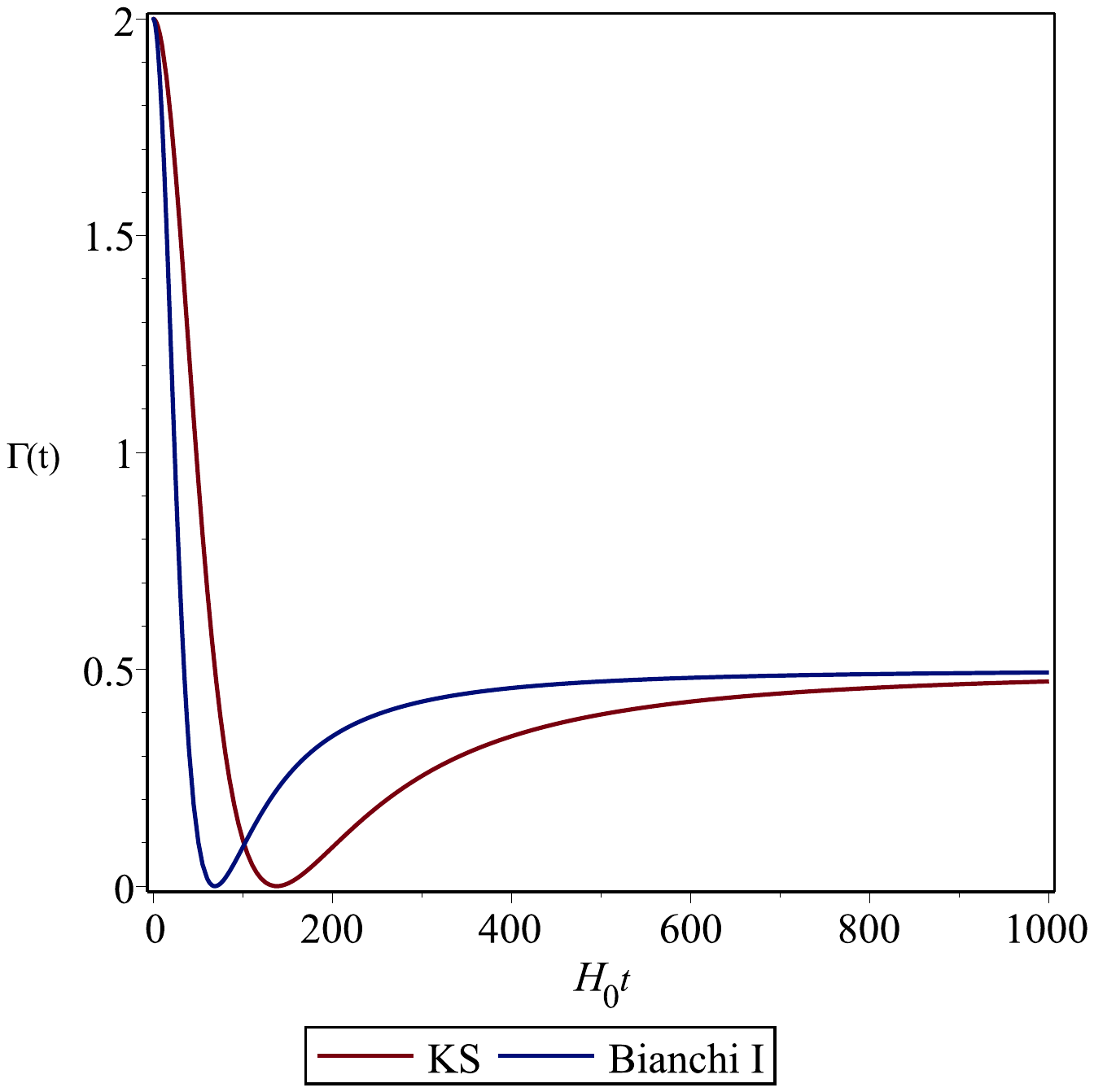}~\hspace*{-3.5cm}
	\vspace*{-5cm}
	\caption{variation of anisotropy parameter, $\Gamma(t)$ with time, $H_0t$} \label{fig:case7}
\end{figure}

On integrating the above equation we get
\begin{equation}
	B=\sqrt{\alpha_1t^2+\frac{\alpha_2}{\alpha_1}}
\end{equation} and
\begin{equation}
	A=\frac{\frac{\alpha_1 t}{K}}{\sqrt{\alpha_1 t^2+\frac{\alpha_2}{\alpha_1}}}.
\end{equation}
In this case we obtain physically acceptable solutions even  if $\alpha_2=0$ where $B$ varies linearly with $t$ and $A$ is a constant.
The anisotropy parameter is given by,
\begin{equation}
	\Gamma=\left[\frac{\frac{2}{t^2} + \frac{8t^2\alpha_1^2}{\left(\alpha_1 t^2 + \frac{\alpha_2}{\alpha_1}\right)^2} - \frac{8\alpha_1}{\left(\alpha_1 t^2 + \frac{\alpha_2}{\alpha_1}\right)}}{\frac{1}{t^2} + \frac{t^2\alpha_1^2}{\left(\alpha_1 t^2 + \frac{\alpha_2}{\alpha_1}\right)^2} + \frac{2\alpha_1}{\left(\alpha_1 t^2 + \frac{\alpha_2}{\alpha_1}\right)}}\right]^2,
\end{equation}
The variation of the anisotropy with time is shown in Fig(\ref{fig:case7}). The initial anisotropy washes out but in future the universe will transit to an anisotropic one.  

The perturbation is given by

\begin{equation}
	\epsilon(t)=\frac{Ct^2 e^{-t}}{\alpha_1 t^2+\frac{\alpha_2}{\alpha_1}}.
\end{equation}

The perturbation over time evolves as shown in Figure 8.

\begin{figure}[t]
	\centering
	\includegraphics[scale=0.4]{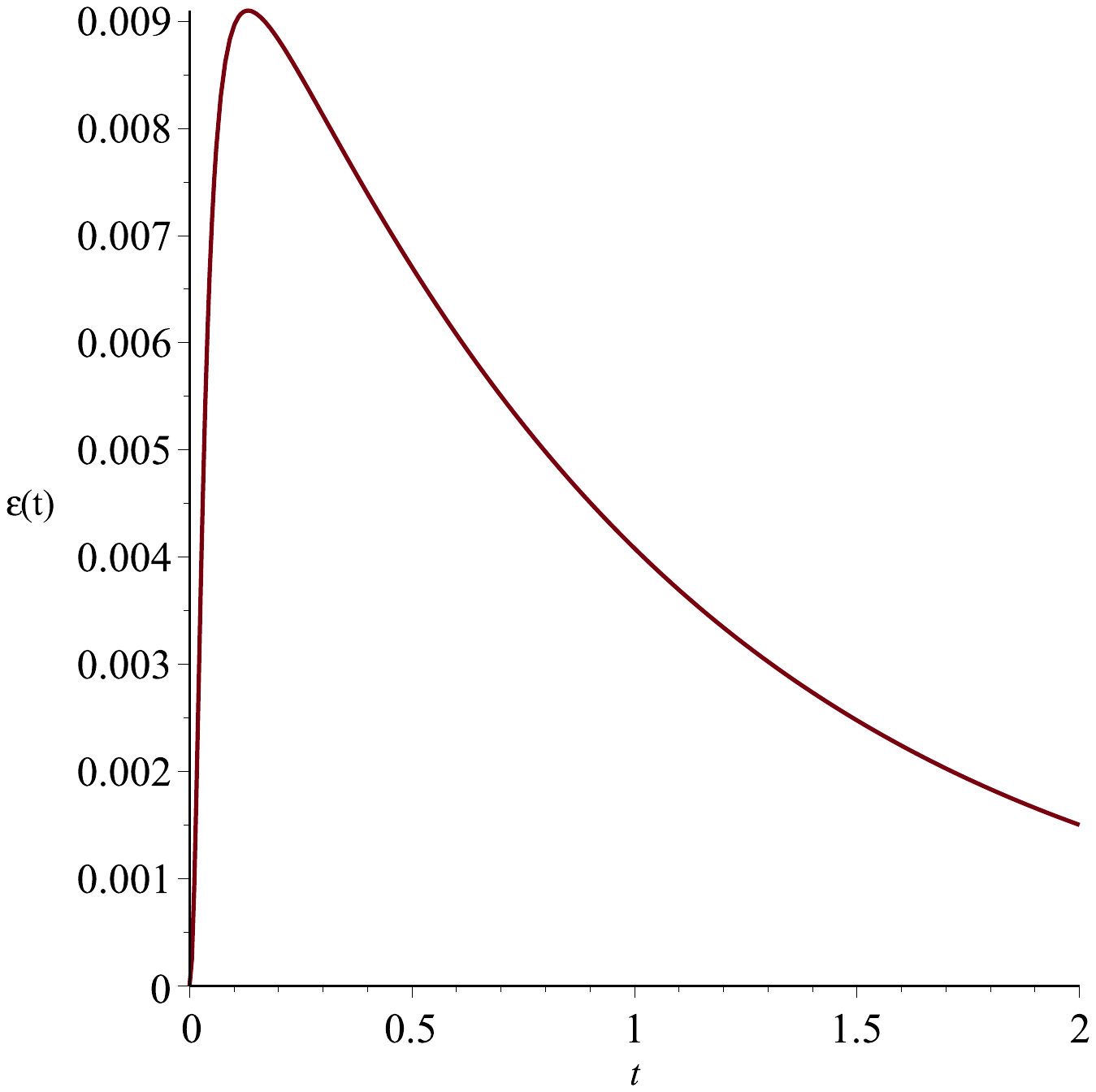}~\hspace*{-3.5cm}
	\vspace*{-5cm}
	\caption{variation of perturbation, $\epsilon(t)$ with time, $t$} \label{fig:case8}
\end{figure}

\subsection{\bf Bianchi-I Model-II}

In this model we consider $\frac{1}{B}=H_0$ which implies  $B=H_0^{-1}$.
From the first two EFE we determine
\begin{equation}
	\Lambda=-\alpha_1H_0^2-\alpha_2H_0^4.
\end{equation}
The coupling parameters $\kappa$ and $\lambda$ may be  positive or zero which leads to both  $\alpha_1$ and $\alpha_2$ positive quantities. In this case  the cosmological constant may be considered $negative$ for a physically viable  cosmological solutions.  The EFE given by eq. (43) becomes
\begin{equation}
	\frac{\ddot{A}}{A}= - 2 \alpha_2 \; H_0^2 \; \left[\frac{\alpha_1}{2\alpha_2}+H_0^2\right].
\end{equation}
which admits an  $oscillatory$ universe which is new and interesting.

\section{Discussion}

Two different anisotropic universes namely,  Kantowski-Sachs and Bianchi-I metrics are considered in the presence of a cosmological constant with Skyrme fluid as the matter source. In general, for all types of solutions the parameters associated with the Skyrme fluid are playing important role in determining the evolution of the universe. Except for Case-B of the KS universe we note existence of a universe with a large initial anisotropy which however, is found to isotropize at a later epoch. The isotropization of the anisotropic universe possible in the absence of any inflationary mechanism involving the scalar inflaton field and even in the absence of the cosmological constant. This feature suggest that Skyrme fluid as a source of matter helps in isotropizing the anisotropic universe leading to  an accelerationg expansion at the present epoch. We also note that an anisotropic  universe transits to the observed isotropic and homogeneous universe at the present epoch even with or without  $\Lambda$ in the presence of Skyrme fluid. 

The behaviour of the anisotropic parameter in the KS and Bianchi-I universes are plotted in Figs. 3, 4, 6 and 7 for different cases. The plots of anisotropic parameter in Fig. 3 for  both KS and Bianchi-I universes show that the  initial anisotropy may be high for the former universe but eventually it washes out at almost same values of $H_0 t$, thus indicating a significantly higher isotropization rate for  a KS universe  compared to that of a Bianchi-I universe. In Fig.4,  we draw the anisotropic parameter for both the  the universes corresponding to Case A and found that  KS universe  is always anisotropic but  Bianchi-I universe  isotropizes. 

It is noted that when Skyrme fluid dominates  in  KS and Bianchi-I universes separately, it permits a cosmology with initial large anisotropy which at a later epoch transforms to isotropic universe. A new solution in Bianchi-I universe is found here which permits an ${emergent}$ universe for a negative $\Lambda$ which, however, cannot be realized in  the KS universe. The behaviour of the anisotropic parameters for both the universes  are plotted in Fig-6. It is evident  that the initial anisotropy is same for both the universes which eventually drops down to zero at a later epoch but the rate of isotropization is faster for the Bianchi-I universe compared to the KS-universe. 

In the absence of  cosmological constant ($\Lambda$), the  initial large anisotropy decreases  for both the universes but it transits to a universe  which once again will transit to an anisotropic universe at a later stage of evolution. This is possibly due to the absence of the repulsive effect of the $\Lambda$ term, but physically it may be due  to the structure formation. The initial anisotropy drops down faster and returns earlier for Bianchi-I universe,while isotropization is  comparatively slow for KS universe and further growth of anisotropy is at a considerably slower rate.  In this case KS geometry gives no well-behaved acceptable solution but a Bianchi-I geometry admits an acceptable solution.

In the case of  $Model-II$ cosmological solutions for KS and Bianchi-I space-times sourced by a Skyrme fluid we noticed an interesting feature which has a sharp contrast between the two anisotropic universe.
 A KS universe with Skyrme fluid is found to admit  three types of cosmological solutions, namely,  linear, expanding and oscillatory  for  different constraints on $H_0^2$, but in the case of Bianchi I space-time, it permits only $oscillatory$ solution. 

We consider in detail two particular solutions where we have considered the KS and Bianchi-I universes to be filled with the Skyrme fluid, without any contribution from the cosmological constant $\Lambda$, which has been set to zero (Case-D). We determine the  two physical parameters, namely the deceleration and the jerk parameters for both the KS and Bianchi-I universes in terms of the redshift parameter $z$ and plotted  the variation as a function of $z$ in order to see how well they confront with observational data. If they can represent the data well, it implies that the initially anisotropic universe described by KS and Bianchi I spacetimes, sourced by a Skyrme fluid only, not only isotropizes at late times but also can describe the observational data well enough at low redshifts. 

In the KS universe sourced only by the Skyrme fluid, the deceleration parameter $q$ defined as $q=-\frac{a\ddot{a}}{\dot{a}^2}$, where $a=(AB^2)^\frac{1}{3}$ gives the average scale factor, is obtained to have the form

\begin{equation}
	q=\frac{\frac{10(z+1)^2}{9f}-\frac{(\alpha_1-1)^3(z+1)^8f}{9K^2}+\frac{7(\alpha_1-1)^6(z+1)^{14}f^3}{9K^4}}{\frac{(z+1)^2}{9f}+\frac{2(\alpha_1-1)^3(z+1)^8f}{9K^2}+\frac{(\alpha_1-1)^6(z+1)^{14}f^3}{9K^4}}-2
\end{equation}
where we define the parameter $f$ as
\begin{equation}
	f=\frac{\frac{\alpha_2}{\alpha_1-1}+\sqrt{\left(\frac{\alpha_2}{\alpha_1-1}\right)^2+\frac{4K^2}{(z+1)^6(\alpha_1-1)}}}{2(\alpha_1-1)}.
\end{equation}
The jerk parameter $j$ defined as $j=\frac{\frac{d^3 B}{dt^3}}{aH^3}$, is calculated for the KS universe as
\begin{equation}
	j(z)=\frac{\frac{d^3 z}{dt^3}}{{\dot{z}}^3}(1+z)^2-\frac{6\ddot{z}(1+z)}{{\dot{z}}^3}+6
\end{equation}
where,
$\dot{z}=-\frac{(\alpha_1-1)^3(z+1)^7A^\frac{3}{2}}{3K^2}-\frac{(z+1)A^\frac{-1}{2}}{3}$

$\ddot{z}=-\frac{7(\alpha_1-1)^3(z+1)^6 \dot{z} A^\frac{3}{2}}{3K^2}-\frac{\dot{z}A^\frac{-1}{2}}{3}+\frac{(z+1)A^{-1}}{3}$

$\frac{d^3 z}{dt^3}=-\frac{14(\alpha_1-1)^3(z+1)^6 \dot{z}A}{K^2}-\frac{2(\alpha_1-1)^3(z+1)^6A^\frac{1}{2}}{K^2}-\frac{14(\alpha_1-1)^3(z+1)^5 {\dot{z}}^2A^\frac{3}{2}}{K^2}-\\ ~~~\frac{7(\alpha_1-1)^3(z+1)^6 {\ddot{z}}A^\frac{3}{2}}{K^2}-\frac{\ddot{z}A^\frac{-1}{2}}{3}+\frac{2\dot{z}A^{-1}}{3}-\frac{2(z+1)A^\frac{-3}{2}}{3}$,

$A=\frac{\frac{\alpha_2}{\alpha_1-1}+\sqrt{\left(\frac{\alpha_2}{\alpha_1-1}\right)^2+\frac{4K^2}{(\alpha_1-1)(z+1)^6}}}{2(\alpha_1-1)}$. \\

In  the Bianchi I universe, the decleration parameter takes the form
\begin{equation}
	q=\frac{\frac{10(z+1)^2}{9f}-\frac{\alpha_1^3(z+1)^8f}{9K^2}+\frac{7\alpha_1^6(z+1)^{14}f^3}{9K^4}}{\frac{(z+1)^2}{9f}+\frac{2\alpha_1^3(z+1)^8f}{9K^2}+\frac{\alpha_1^6(z+1)^{14}f^3}{9K^4}}-2
\end{equation}
where
\begin{equation}
	f=\frac{\frac{\alpha_2}{\alpha_1}+\sqrt{\left(\frac{\alpha_2}{\alpha_1}\right)^2+\frac{4K^2}{(z+1)^6\alpha_1}}}{2\alpha_1}
\end{equation}
and the jerk parameter is computed which is given by
\begin{equation}
	j(z)=\frac{\frac{d^3 z}{dt^3}}{{\dot{z}}^3}(1+z)^2-\frac{6\ddot{z}(1+z)}{{\dot{z}}^3}+6
\end{equation}
where,
$\dot{z}=-\frac{\alpha_1^3(z+1)^7A^\frac{3}{2}}{3K^2}-\frac{(z+1)A^\frac{-1}{2}}{3}$

$\ddot{z}=-\frac{7\alpha_1^3(z+1)^6 \dot{z} A^\frac{3}{2}}{3K^2}-\frac{\dot{z}A^\frac{-1}{2}}{3}+\frac{(z+1)A^{-1}}{3}$

$\frac{d^3 z}{dt^3}=-\frac{14\alpha_1^3(z+1)^6 \dot{z}A}{K^2}-\frac{2\alpha_1^3(z+1)^6A^\frac{1}{2}}{K^2}-\frac{14\alpha_1^3(z+1)^5 {\dot{z}}^2A^\frac{3}{2}}{K^2}-\\ ~~~\frac{7\alpha_1^3(z+1)^6 {\ddot{z}}A^\frac{3}{2}}{K^2}-\frac{\ddot{z}A^\frac{-1}{2}}{3}+\frac{2\dot{z}A^{-1}}{3}-\frac{2(z+1)A^\frac{-3}{2}}{3}$,

$A=\frac{\frac{\alpha_2}{\alpha_1-1}+\sqrt{\left(\frac{\alpha_2}{\alpha_1-1}\right)^2+\frac{4K^2}{\alpha_1(z+1)^6}}}{2\alpha_1}$.\\

We plot  $q$ and $j$ parameters as functions of the redshift parameter $z$ in Figs. (9) and (10) respectively.
For the plot we have taken 
$K = 0.02$, $\alpha_1= 0.5$,  $\alpha_2 := .5\times10^{-31}$. All the chosen parametric values are well within the allowed ranges as mentioned in the introductory section.
\begin{figure}[t]
	\centering
	\includegraphics[scale=0.4]{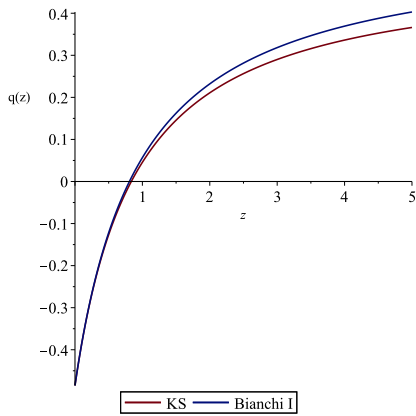}~\hspace*{-3.5cm}
	\vspace*{-5cm}
	\caption{variation of deceleration parameter, $q(z)$ with $z$} \label{fig:case9}
\end{figure}

\begin{figure}[t]
	\centering
	\includegraphics[scale=0.4]{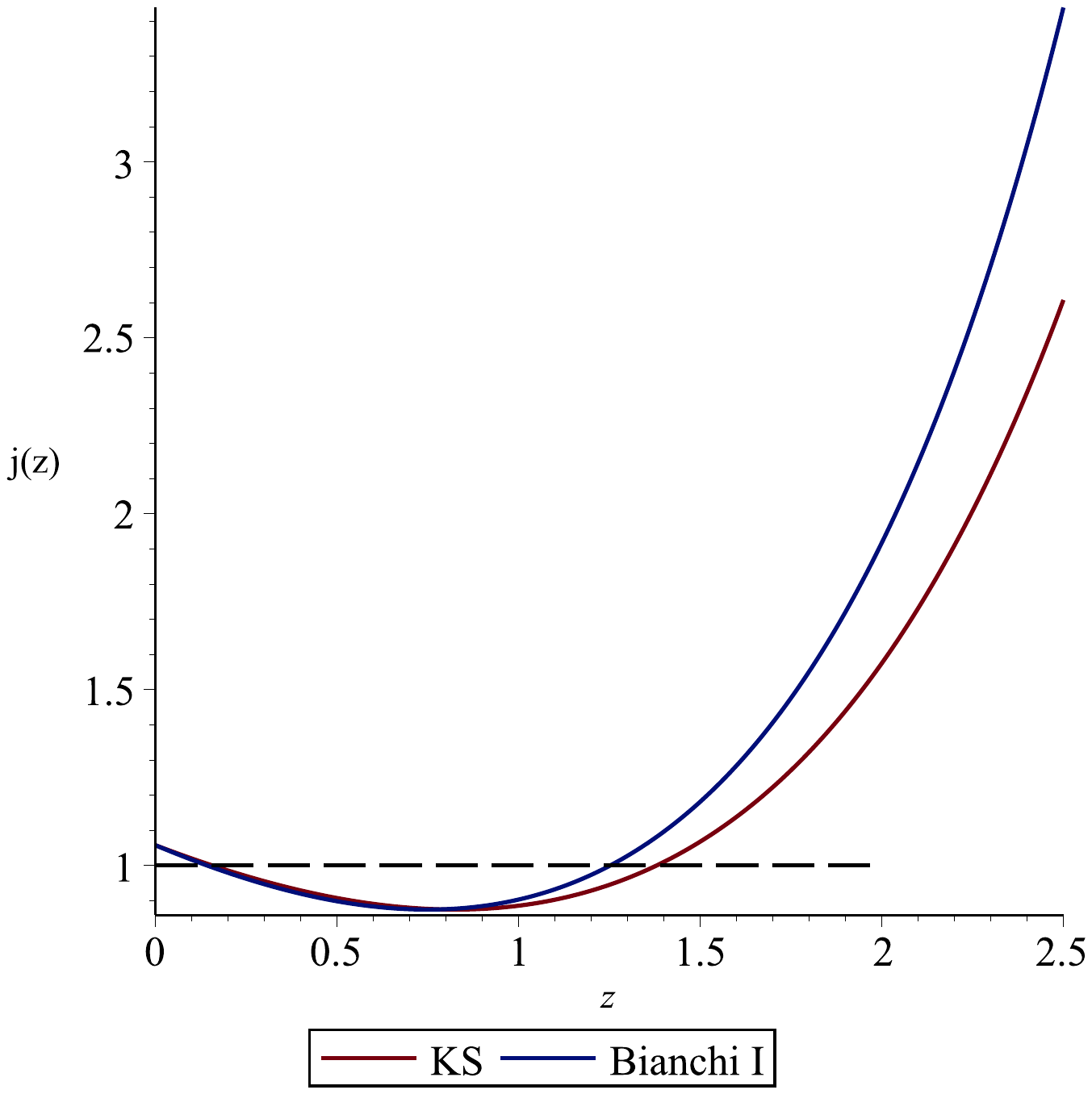}~\hspace*{-3.5cm}
	\vspace*{-5cm}
	\caption{variation of jerk parameter, $j(z)$ with $z$} \label{fig:case10}
\end{figure}

A number of interesting conclusions can be drawn from the plots. As we can see from the plot for $q$, the deceleration parameter takes an approximate value $-0.55$ at the present times as the redshift parameter approaches towards zero value, which is well in agreement with observational data\cite{Bamba}. The jerk parameter becomes unity for very small redshift which corresponds well with the $\Lambda$CDM model. Also, there is a sign flip that we observe in the deceleration parameter from positive to negative which physically represents the transition from declerated to accelerated expansion in the present times at a redshift value $z \simeq 0.8$, which is again in good agreement with observations \cite{Bamba}. If we extrapolate the plot to even higher redshifts, it is evident that $q$ tends to 0.5, which  agrees well with the predicted value \cite{Bamba}. 

We  obtained here  the time-dependent perturbation introduced in the field equations.  The temporal variation of the perturbation can be plotted but for convenience we plot only those $\epsilon$ which are of physical interest. In the Case-D of KS universe and Case-A, B and D of the Bianchi-I universe, the evolution of the perturbation with time relates to the physical universe. In these cases the small perturbation  either begins at a small finite maximum and then decreases monotonically before appproaching a very small value in near approximity to zero. However, in the  case-D of Bianchi universe, the perturbation increases slightly before reaching a maxima at the early time and then falling off. All these four instances where the perturbation falls of with time indicates stable solutions. In the other cases the perturbation either increases with time or remains constant which does not permit stability of the cosmological models. 

A very interesting observation is that, at the present epoch with redshift parameter  less than 1, both the KS and Bianchi-I universes containing Skyrme fluid in the $absence$ of a $\Lambda$ term replicates the behaviour of the standard $\Lambda$CDM model along with the decleration parameter which corresponds very well with observations. Also, as we can see from the perturbative stability analysis, both the cases for the KS and Bianchi-I universes in the absence of the $\Lambda$ term are characterized by stable solutions. For higher redshift values the behaviour of the two universes are not exactly identical but still in close proximity such that no contrasting physical feature is obtained as in Model- II. Thus we can conclude that both the universes, initially anisotropic, filled with Skyrme fluid in the absence of a cosmological constant term isotropizes over time and  corresponds well with the observational data at the present time, also satisfying some essential features of the standard $\Lambda$CDM model. This justifies the physical motivation for our work. Thus, we claim Skyrme fluid as a possible source which  isotropizes the anisotropic universe  $ via$  the present accelerated expansion, also capable of reproducing certain features of the present observerd universe. However, the structure formation process in such a scenario has not been dealt with in the present work and we plan to do so in the near future.

\section{APPENDIX}

EFE for vacuum with non-zero $\Lambda$: $R_\nu^\mu=\Lambda \delta_\nu^\mu$.

$Computation$ $of$ $Ricci$ $Tensor$

\begin{equation}
	R_0^0=\Lambda=\ddot{(ln V)}+H_1^2+H_2^2+H_3^2=\frac{\ddot{R_1}}{R_1}+\frac{\ddot{R_2}}{R_2}+\frac{\ddot{R_3}}{R_3}.
\end{equation}

\begin{equation}
	R_1^1=\Lambda=\frac{\dot{(VH_1)}}{V}=\frac{\dot{R_1}\dot{R_2}}{R_1 R_2}+\frac{\dot{R_1}\dot{R_3}}{R_1 R_3}+\frac{\ddot{R_1}}{R_1}.
\end{equation}

\begin{equation}
	R_2^2=\Lambda=\frac{\dot{(VH_2)}}{V}=\frac{\dot{R_2}\dot{R_3}}{R_2 R_3}+\frac{\dot{R_1}\dot{R_2}}{R_1 R_2}+\frac{\ddot{R_2}}{R_2}.
\end{equation}

\begin{equation}
	R_3^3=\Lambda=\frac{\dot{(VH_3)}}{V}=\frac{\dot{R_2}\dot{R_3}}{R_2 R_3}+\frac{\dot{R_1}\dot{R_3}}{R_1 R_3}+\frac{\ddot{R_3}}{R_3}.
\end{equation}

$Computation$ $of$ $Ricci$ $Scalar$

\begin{equation}
	R=2[\frac{\ddot{R_1}}{R_1}+\frac{\ddot{R_2}}{R_2}+\frac{\ddot{R_3}}{R_3}+\frac{\dot{R_1}\dot{R_2}}{R_1 R_2}+\frac{\dot{R_1}\dot{R_3}}{R_1 R_3}+\frac{\dot{R_2}\dot{R_3}}{R_2 R_3}].
\end{equation}

$Computation$ $of$ $Einstein$ $Tensor$

\begin{equation}
	G_0^0=-\frac{\dot{R_1}\dot{R_2}}{R_1 R_2}-\frac{\dot{R_1}\dot{R_3}}{R_1 R_3}-\frac{\dot{R_2}\dot{R_3}}{R_2 R_3}.
\end{equation}

\begin{equation}
	G_1^1=-\frac{\ddot{R_2}}{R_2}-\frac{\ddot{R_3}}{R_3}-\frac{\dot{R_2}\dot{R_3}}{R_2 R_3}.
\end{equation}

\begin{equation}
	G_2^2=-\frac{\ddot{R_1}}{R_1}-\frac{\ddot{R_3}}{R_3}-\frac{\dot{R_1}\dot{R_3}}{R_1 R_3}.
\end{equation}

\begin{equation}
	G_3^3=-\frac{\ddot{R_1}}{R_1}-\frac{\ddot{R_2}}{R_2}-\frac{\dot{R_1}\dot{R_2}}{R_1 R_2}.
\end{equation}

Putting $R_1=A$ and $R_2=R_3=B$, we obtain Equations(38)-(40).


\begin{thebibliography}{99}
	
	
	\bibitem{Misner} C. W. Misner, Ap. J. \textbf{151}, (1968) 431.	
	
	\bibitem{Huang}  W. Huang, J. Math. Phys. \textbf{31}, (1990) 1456.
	
	\bibitem{Chimento} L. P. Chimento et al, Class. Quantum Gravit. \textbf{14}, (1997) 3363.
	
	\bibitem{Saha1} B. Saha, Int. J. Theor. Phys. \textbf{45}, (2006) 983.
	
	\bibitem{Saha2}  B. Saha, Astrophys. Space Sci. \textbf{302}, (2006) 83.
	
	\bibitem{Pradhan1} A. Pradhan and P. Pandey, Astrophys. Space Sci. \textbf{301}, (2006) 221.
	
	\bibitem{Pradhan2} A. Pradhan and S. K. Singh, Int. J. Mod. Phys. \textbf{D13}, (2004) 503.
	
	\bibitem{Bianchi} L. Bianchi, Mem. di Mat. e di Fis. della Soc. Ita. delle Sci., \textbf{11} (1898) 267.
	
	\bibitem{Taub} A. H. Taub, Ann. of Math. (1951) 472.
	
	\bibitem{Heckmann} O. Heckman, and E. Schücking, edited by L. Witten (1962) 5525.
	
	\bibitem{Ellis} G. F. R Ellis and M. A. H. MacCallum, Comm. in Math. Phys. \textbf{12} (1969) 108.
	
	\bibitem{Kantowski} R. Kanotowski and R.K. Sachs, J.Math.Phys. \textbf{7} (1966) 443.
	
	\bibitem{Collins} C. B. Collins, J.Math.Phys. \textbf{18} (1977) 2116.
	
	\bibitem{Weber} E. Weber, J.Math.Phys. \textbf{25} (1984) 3279.
	
	\bibitem{Linde} A. D. Linde, Phys.Lett. \textbf{162B} (1985) 281.
	
	\bibitem{Paul} B. C. Paul, D. P. Datta and S. Mukherjee, Mod. Phys. Lett. A \textbf{01} (1986) 149.
	
	\bibitem{Dunsby1} M. Bradley, P. K. S. Dunsby, M. Forsberg and Z. Keresztes, Class. Quant. Grav. \textbf{29} (2012) 095023
	
	\bibitem{Dunsby2} M. Bradley, M. Forsberg, Z. Keresztes, L. . Gergely and P. K. S. Dunsby, arXiv:1303.4576 [gr-qc].
	
	\bibitem{Dunsby3} Z. Keresztes, M. Forsberg, M. Bradley, P. K. S. Dunsby and L. . Gergely, Springer. Pro. in Math. and Stat., (2014) 289 .
	
	\bibitem{Canfora} F. Canfora and H. Maeda, Phys. Rev. D \textbf{87} (2013) 084049
	
	\bibitem{Parisi} L. Parisi, N. Radicella, G. Vilasi Phys.Rev. D \textbf{91} (2015) 063533.
	
	\bibitem{Skyrme} T. Skyrme, Nucl. Phys., \textbf{31} (1962) 556.
	
	\bibitem{Kalbermann} G. Kalbermann, Nucl.Phys. A \textbf{612} (1997) 359.
	
	\bibitem{Khawaja} U. Al Khawaja, H. Stoof, Nature. \textbf{411} (2001) 918.
	
	\bibitem{Baskaran} G. Baskaran, arXiv:1108.3562 (2011).
	
	\bibitem{Canfora2} F. Canfora, Phys. Rev. D \textbf{88} (2013) 065028.
	
	\bibitem{Mukherjee} S. Mukherjee, B. C. Paul, N. K. Dadhich, S. D. Maharaj, and A. Beesham, Class.  and Quant. Grav. \textbf{23} (2006) 6927.
	
	\bibitem{Banerjee} A. Banerjee, T. Bandyopadhyay, and S. Chakraborty, Gen.Rel.and Grav.\textbf{40} (2008) 1603.
	
	\bibitem{Bamba} A. A. Mamon, K. Bamba, Eur. Phys. J. C \textbf{78} (2018) 862. 
	
	
\end{thebibliography}
\end{document}